\documentstyle[11pt]{article}

\def \n {\noindent}
\def\R{\ifmmode{\rm I\mkern-3.1mu
R\mkern1mu}\else{\rm I\kern-.18em
R\hskip1pt\ }\fi\relax}

\def\Z{\ifmmode{\it Z\mkern-7.8mu
Z\mkern2mu}\else{\it Z\kern-.28em
Z\hskip1pt\ }\fi\relax}

\begin{document}



\newtheorem{theoreme}{{\bf Th\'eor\`eme}}[section]
\newtheorem{corollaire}{{\bf Corollaire}}[section]
\newtheorem{lemme}{{\bf Lemme}}[section]
\newtheorem{proposition}{{\bf Proposition}}[section]
\newtheorem{definition}{{\bf Definition}}[section]
\newtheorem{remarque}{{\bf Remarque}}[section]

\newcommand{\Lim}{\mathop{\hbox{\rm Lim}}}

\def\buildrel#1\over#2{\mathrel{\mathop{\kern 0pt#2}\limits^{#1}}}
\def\build#1_#2^#3{\mathrel{\mathop{\kern 0pt#1}\limits_{#2}^{#3}}}

\def\diagram#1{\def\normalbaselines{\baselineskip=0pt
\lineskip=10pt\lineskiplimit=1pt}  \matrix{#1}}

\def\hfl#1#2{\smash{\mathop{\hbox to 12mm{\rightarrowfill}}
\limits^{\scriptstyle#1}_{\scriptstyle#2}}}

\def\vfl#1#2{\llap{$\scriptstyle #1$}\left\downarrow
\vbox to 6mm{}\right.\rlap{$\scriptstyle #2$}}

\def\toto{\leaders \hbox to 4mm{\hfil.\hfil}\hfill}

\newcommand{\Log}{{\hbox{{\rm Log}}}}
\newcommand{\Sup}{{\hbox{{\rm Sup}}}}
\newcommand{\Inf}{{\hbox{{\rm Inf}}}}
\def\Supp_#1{\mathop{\hbox{\rm Sup}}\limits_{#1}}
\def\Inff_#1{\mathop{\hbox{\rm Inf}}\limits_{#1}}

\begin{center}

  {\Large \bf Regularized trace formula of magic Gribov operator\\ on Bargmann space}\\

\end{center}
\begin{center}
 by
\end{center}
\begin{center}
 {\it Abdelkader Intissar $^{(*),(**)}$}

 {\it $^{(*)}$ Equipe d'Analyse spectrale , Facult\'e des Sciences et Techniques, Universit\'e
de Cort\'e, 20250 Cort\'e, France}\\
{\it T\'el: 00 33 (0) 4 95 45 00 33}\\
{\it Fax: 00 33 (0) 4 95 45 00 33}\\
{\it intissar@univ-corse.fr}\\
{\it $^{(**)}$ Le Prador, 129, rue du Commandant Rolland, 13008 Marseille,
France}\\
\end{center}
\begin{center}
*****
\end {center}
\abstract{ {\it In this article, we obtain a regularized trace formula for magic Gribov operator\\ $ H = \lambda{''}G + H_{\mu,\lambda}$ acting on Bargmann space where $$G = a^{*3}a^{3} \quad \quad and \quad \quad  H_{\mu,\lambda} = \mu a^{*}a + i\lambda a^{*}( a + a^{*})a$$ Here $a$ and $a^{*}$ are the standard Bose annihilation and creation operators and in Reggeon field theory, the real parameters $\lambda{''}$ is the magic coupling of Pomeron, $\mu$  is Pomeron intercept, $\lambda$ is the triple coupling of Pomeron and $i^{2} = -1$.\\

An exact relation is established between the degree of subordination of the perturbation operator $H_{\mu,\lambda}$ to the unperturbed operator $G$ and the number of corrections necessary for the existence of finite formula of the trace.}}\\


 \begin{center}
 {\bf 1. Introduction and main result}
 \end{center}
 As is known, the trace of a finite-dimensional matrix is the sum of all the eigenvalues. But in an infinite dimensional space, in general, ordinary differential operators do not have a finite trace.\\
 In 1953, Gelfand and Levitan considered the Sturm-Liouville operator\\

$ \cases{-y''(x) + q(x)y(x) = \sigma y(x)\cr y'(0) = 0, y'(\pi) = 0\cr q(x) \in¸ C^{1} [0, \pi],\quad \displaystyle{\int_{0}^{\pi}q(x)dx} = 0\cr}$ $\hfill { }  (*)$\\

and derived the formula\\

$\displaystyle{\sum_{n=1}^{\infty}(\sigma_{n} -\lambda_{n}) = \frac{1}{4}(q(0) + q(\pi))}$ $\hfill { }  (**)$\\

where $\sigma_{n}$ are the eigenvalues of the above operator and $\lambda_{n} = n^{2}$ are the eigenvalues of the same operator with $q (x) = 0$.\\

The proof of this regularized trace formula for the Sturm-Liouville operator can been found in {\bf[6]}.\\

The same regularized trace formula for the same problem was obtained with different method by Dikii {\bf[4]}.\\

 For the scalar Sturm-Liouville problems, there is an enormous literature (see for example {\bf[5]} or {\bf[21]}) on estimates of large eigenvalues and regularized trace formulae which may often be computed explicitly in terms of the coefficients of operators and boundary conditions.\\

After these studies, several mathematicians were interested in developing regularized trace formulae for different differential operators. According Sadovnichii and  Podol'skii, these formulae gave rise to a large and very important theory, which started from the investigation of specific operators and further embraced the analysis of regularized traces of discrete operators in general form.\\

Among the results of Sadovnichii and  Podol'skii established for abstract operators, we can recall that following :\\

Let $A_{0}$ be a self-adjoint positive discrete operator of domain $D(A_{0})$ acting in a Hilbert space, $\{\lambda_{n}\}$ be its eigenvalues arranged in ascending order, $\{\phi_{n}\}$ be a basis formed by the eigenvectors of $A_{0}$, $B$ be a perturbation operator, and $\{\sigma_{n}\}$ be the eigenvalues of $A_{0} + B$. Also, assume that $A_{0}^{-1}$ is a trace class operator.\\

For operators $A_{0}$ and $B$ in ({\bf[22]}, Theorem 1), the following theorem is proved.\\

{\bf Theorem 1.1.}
{\it Let the operator $B$ be such that $D(A_{0}) \subset D(B)$, and let there exists a number $\delta \in [0, 1)$ such that $BA_{0}^{-\delta}$ has a bounded extension, and number $\omega \in [0, 1)$, $ \omega + \delta < 1$ such that $A_{0}^{-(1-\delta -\omega)}$ is a trace class operator.\\ Then, there exist a subsequence of natural numbers $\{n_{m}\}_{m=1}^{\infty}$and a subsequence of contours $\Gamma_{m} \in I\!\!\!\! C$, that for $ \omega \geq \delta $ the formula\\

$lim \displaystyle{\sum_{j=1}^{n_{m}}(\sigma_{j} - \lambda_{j} - \mid <B\phi_{j}, \phi_{j}>\mid) = 0}$ $m\rightarrow \infty$ is true.\\

{\bf Remark 1.2.}
{\it 1) This theorem has been successfully applied to concrete ordinary differential operators as well as to partial differential operators, but some of its assumptions are not verified for our specific operator that we shall study.\\
2) We can found in {\bf[23]} or in {\bf[24]}, an excellent survey dedicated by Sadovnichii and  Podol'skii to the history of the state of the art in the theory of regularized traces of linear differential operators with discrete spectrum and a detailed list of publications related to the present aspect.}$\hfill { }  \Box $\\

Usually, quantum Hamiltonians are constructed as self-adjoint operators; for certain situations,
however, non-self-adjoint Hamiltonians are also of importance. In particular, the reggeon field
theory (as invented by V. Gribov {\bf[9]}) for the high energy behaviour of soft processes is governed by the magic non-self-adjoint Gribov operator\\

$H_{\lambda'',\lambda',\mu,\lambda} = \lambda{''}a^{*3}a^{3}+\lambda{'}a^{*2}a^{2} + \mu a^{*}a +i\lambda a^{*}(a + a^{*})a $ $\hfill {(1.1) } $\\

\n where $a$ and its adjoint $a^{*}$ are annihilation and creation operators, respectively, satisfying the canonical commutation relation $[a, a^{*}] = I$.\\

In the case $\lambda{''} = 0$, Ando and Zerner in {\bf[2]} and Intissar in {\bf[10 - 14]} have given a complete spectral theory for the operator $H_{0,\lambda{'},\mu,\lambda}$.\\
 In particular, one can consult the list of spectral properties of this operator summarized in ({\bf[15]}, theorm 1.2, p.671).\\

In the case $\lambda{''} \neq  0$ and $\lambda{'} = 0$, Aimar et al in {\bf[1]} have given some spectral properties of magic Gribov operator $H_{\lambda{''},0,\mu,\lambda}$, which is more regular than $H_{0,\lambda{'},\mu,\lambda}$.\\

In this paper, we continue the spectral study of $H_{\lambda{''},0,\mu,\lambda}$ to obtain a regularized trace formula of this operator.\\

We adopt the following notations: \\

$ H = H_{\lambda{''},0,\mu,\lambda} = \lambda{''}G + H_{\mu,\lambda}$
where
$G = a^{*3}a^{3}$ \quad and \quad $H_{\mu,\lambda} = \mu a^{*}a + i\lambda a^{*}( a + a^{*})a$ \\ Here $a$ and $a^{*}$ are the standard Bose annihilation and creation operators and $\lambda{''}$, $\mu$, $\lambda$ are real parameters and $i^{2} = -1$.\\

It is convenient to regard the above operators as acting on Bargmann space $E$ {\bf[3]}.\\

$E$ is defined as a subspace of the space O($ (I\!\!\!\!C$) of holomorphic functions on $I\!\!\!\!C$\, given by\\

$ E = \{\phi\in O(I\!\!\!\!C) ; < \phi, \phi > < \infty \} $ $\hfill {(1.2) } $\\

where the paring \\

$<\phi,\psi> = \displaystyle{\int_{I\!\!\!\!C}}\displaystyle{\phi(z)\overline{\psi(z)}e^{-\mid z\mid^{2}}dxdy}$ $\forall$ $\phi,\psi \in O(I\!\!\!\!C)$ $\hfill {(1.3) } $\\

 and $dxdy$ is Lebesgue measure on $I\!\!\!\!C$.\\

The Bargmann space $E$ with $\mid\mid \phi \mid\mid = \sqrt{<\phi, \phi>}$ is a Hilbert space and $e_{n}(z) = \frac{z^n}{\sqrt{n!}}; n = 0, 1, ....$ is an orthonormal basis in $E$.\\

In this representation, the standard Bose annihilation and creation operators are defined by\\
\begin{flushleft}
$ \cases{a\phi(z) = \quad \phi^{'}(z)\cr with \quad maximal \quad domain \cr
D(a) = \{\phi \in E \quad such \quad that \quad  a\phi \in E\} \cr }$ \\
\end{flushleft}
\begin{flushleft}
$ \cases{a^{*}\phi(z) = \quad z\phi(z)\cr with \quad maximal \quad domain \cr
D(a^{*}) = \{\phi \in E \quad such \quad that \quad  a^{*}\phi \in E\} \cr }$ \\
\end{flushleft}
Accordingly, for the operator  $ H := H_{\lambda^{''}, 0,\mu,\lambda}$ we have\\
\begin{flushleft}
$ \cases{H \phi(z)= \lambda^{''} z^{3}\phi^{'''}(z)+i\lambda z\phi^{''}(z) + (i\lambda z^{2} + \mu z)\phi^{'}(z) \quad \quad \quad  \quad \quad \quad \cr with \quad maximal \quad domain \cr
D(H_{max}) = \{\phi \in E ; H\phi \in E\} \cr }$\\
\end{flushleft}

{\bf Remark 1.3.}
{\it Tobin, in an interesting article {\bf[28]}, have derived several formulas of Gelfand-Levitan type for the first regularized trace of discrete operators under various conditions convenient for verification. But, for $\mu \neq 0 $ and $\lambda \neq 0 $ the operator $H_{\mu,\lambda} = H_{0, 0,\mu,\lambda}$ is not in the classes of perturbing linear operators considered by him.\\
Notice that in this case the operator $H_{\mu,\lambda}$ is very far from normal and not only its self-adjoint and skew-adjoint parts do not commute but there is no inclusion in either way between their domains or with the domain of their commutator.\\
It may be noted also that (see {\bf[13]}) if  $\mu = 0 $ and $\lambda \neq 0 $ the spectrum of $H_{\mu,\lambda}$ is $ \sigma(H_{\mu,\lambda}) = I\!\!\!\!C.$}}$\hfill { }  \Box $\\

Let us begin by reviewing the most important properties of the $ H := H_{\lambda^{''},0,\mu,\lambda}$\\

i) Let ${\bf Pol}$ be space of polynomials, then ${\bf Pol}$ is dense in $E$.\\

ii) We define $H_{min}$ as the closure of restriction operator $ H_{\mid_{{\bf Pol}}}$ on the polynomials acting in Bargmann space, i.e.
\begin{flushleft}
$ \cases{H_{min}\phi= \quad \lambda^{''} z^{3}\phi^{'''}(z)+i\lambda z\phi^{''}  + (i\lambda z^{2} + \mu z)\phi^{'} \cr with  \quad domain \cr
D(H_{min}) = \{\phi \in E \quad such that \quad \exists p_{n} \in {\bf Pol}\quad and \quad\psi \in E \quad such that \quad p_{n}\rightarrow \phi\quad and \quad Hp_{n}\rightarrow \psi\}\cr }$\\
\end{flushleft}

iii) $D(H_{min}) = D(H_{max}) = D(G)$ \\

 In Bargmann space, it may be noted also that\\

 iv) The operator $ G = a^{*3}a^{3} $ is positive, self adjoint operator.\\

 v) The functions $e_{n}(z) = \frac{z^n}{\sqrt{n!}}$ are  the orthonormal eigenvectors of $G$ corresponding to the eigenvalues $\lambda_{n} = n(n-1)(n-2)$ for $n \geq 3 $ and $\lambda_{n} = 0 $ for $  n \in \{0, 1,2\}$\\

 vi) Let $\widetilde{G} = G + I$ then  $ <\widetilde{G}\phi, \phi> \geq <\phi, \phi>$  $\forall \phi \in D(G)$\\

vii) $G $ could be replaced by $\widetilde{G}$ or $G + \beta I$ with a scalar $\beta$ without changing the nature of the problem that we will study.\\

viii) in {\bf[1]} (see theorem 3.3, p: 595), Aimar et al have shown that the spectrum of the magic Gribov operator is discrete and that the system of generalized eigenvectors of this operator is an unconditional basis in Bargmann space $E$.$\hfill { }  \Box $\\

The goal of this article consists in establishing news spectral properties of this operator, in particular to establish an exact relation between the degree of subordination of the non-self-adjoint perturbation operator $H_{\mu,\lambda}$ to the unperturbed operator $G$ and the number of corrections necessary for the existence of finite formula of the regularized trace.\\

Then the main results to which is aimed this paper can be stated as follows for the magic Gribov operator \\

{\bf Theorem 1.4.} {
\it Let $ E $ be the Bargmann space, $ H = \lambda{''}G + H_{\mu,\lambda}$ acting on $E$ where $ G = a^{*3}a^{3}$ and $ H_{\mu,\lambda} = \mu a^{*}a + i\lambda a^{*}( a + a^{*})a$ , $a$ and $a^{*}$ are the standard Bose annihilation and creation operators\\
Then there exists an increasing sequence of radius $r_{n}$ such that $r_{n} \rightarrow \infty$ as $ n \rightarrow \infty$ \\
 and\\
 $\displaystyle{Lim \sum_{k=0}^{n}(\sigma_{k} - \lambda{''}\lambda_{k}) + \frac{1}{2i\pi}\int_{\gamma_{n}} trace[\sum_{j=1}^{4}\frac{(-1)^{j-1}}{j}[H_{\mu,\lambda}(\lambda{''}G - \sigma I)^{-1}]^{j}]d\sigma = 0}$ as $ n \rightarrow \infty$ \\
 Where\\
 - $\sigma_{k}$ are the eigenvalues of the operator $ H = \lambda{''}G + H_{\mu,\lambda}$\\
 - $\lambda_{k} = k(k-1)(k-2)$ are the eigenvalues of the operator $ G $\\
 - $ (\lambda{''}G - \sigma I)^{-1}$ is the resolvent of the operator $\lambda{''}G$\\
 and\\
 - $\gamma_{n}$ is the circle of radius $r_{n}$ centered at zero in complex plane.\\

 We try to be self-contained and elementary as far as possible in this paper which is
organized as follows. In Section 2, we give some properties of subordination of the perturbation operator $H_{\mu,\lambda}$ to the unperturbed operator $G$. In Section 3, we recall introduce the regularized determinant of the perturbation [7] and we give the proof of main results stated in Section 1.\\

 \begin{center}
 {\bf 2. On the degree of subordination of the perturbation operator $H_{\mu,\lambda}$ to the unperturbed operator $G$}
 \end{center}

 We begin this section by improving some basic results of {\bf[1]} on the operator $G$ and by recalling some useful definitions.\\

 For the discreteness of spectrum of the operator $ \tilde{G} = G + I$, it suffices to use the following Rellich's theorem (see {\bf[19]}, p. 386)\\

{\bf Theorem 2.1.}
{\it Let $B$ be a self-adjoint operator in $E$ satisfying $<B\phi, \phi> \geq <\phi, \phi>, \phi \in
D(B)$, where $D(B)$ is a domain of $B$.
Then, the spectrum of $B$ is discrete if and only if the set of all vectors $\phi \in
D(B)$, satisfying $ <B\phi, \phi > \leq 1 $ is a precompact set.}\\

In {\bf[1]}, it was shown the following basic spectral properties on the operator $G$\\

{\bf Lemma 2.2.}{\it 1) The operator $G$ has a compact resolvent.\\
2) Let  $\lambda_{n} = n(n-1)(n-2)$ for $ n > 0 $ the eigenvalues of $G$ associated to eigenvectors $e_{n}(z) = \frac{z^{n}}{\sqrt{n!}}$. Then for each $\sigma \in I\!\!\!\!C$ such that $\sigma \neq \lambda_{n}$, we have
$$(G - \sigma I)^{-1}\phi = \displaystyle{\sum_{n=1}^{\infty}\frac{1}{\lambda_{n} - \sigma}<\phi, e_{n}>e_{n}}.$$\\
Moreover, if $Im \sigma \neq 0$ and if $\sigma$ belongs to a ray with origin zero and of angle $\theta$
with $\theta \neq 0 $ and $\theta \neq \pi $ , we have \\
$$\mid\mid(G - \sigma I)^{-1}\mid\mid \leq \frac{1}{Im \sigma} = \frac{c(\theta)}{\mid \sigma \mid}$$\\
3) There exists a sequence of circles $C(0, r_{n}) , n =1, 2, . . . ,$ with radii $r_{n}$ going to infinity such that\\

$\mid\mid (\widetilde{G} - \sigma I)^{-1}\mid\mid \leq \frac{2}{\mid\sigma\mid^{2/\beta}} $ for any $\beta \geq 3$ and $ \mid \sigma \mid =  r_{n}$ where $ r_{n} = \frac{\lambda_{n} + \lambda_{n+1}}{2}$}\\

{\bf Proof}\\

 1) It is well known that the injection from $D(a)$ into the Bargmann space $E$ is compact and as the injection from  $D(a^{*3}a^{3})$ into $D(a)$ is continuous, then the injection from  $D(a^{*3}a^{3})$ into Bargmann space is also compact. Classically the operators of the form  $ (a^{*3}a^{3} + I) $ are invertible then the resolvent set of $G$ is not void. Consequently the self-adjoint operator $G$ has compact resolvent and this proves again the discreteness of its spectrum.\\

 For the properties 2) and 3) of this lemma see the lemmas 3.1 and 3.2 in {\bf[1]}.$\hfill { }  \Box $\\

 Now, we recall the definition of operators of Carleman's class and some of their properties and we  give a few others spectral properties of the resolvent of the operator $G$.\\

{\bf Definition  2.3.}
{\it A compact operator $K$ belongs to Carleman's class  ${\bf C_{p}}$ of order $p$ if $\displaystyle{\sum_{n=1}^{\infty} s_{n}^{p} < \infty}$, where $s_{n}$ are s-numbers of operator $K$ i.e. the eigenvalues of the operator $\sqrt{K^{*}K}$.\\
In particular, the operator $K$ is called nuclear operator if $K \in {\bf C_{1}}$ and Hilbert-Shmidt operator if $K \in {\bf C_{2}}$.\\
For $ p \geq 1$ the value $(\displaystyle{\sum_{n=1}^{\infty} s_{n}^{p} })^{\frac{1}{p}}$ is a norm on ${\bf C_{p}}$ denoted by $\mid\mid . \mid\mid_{p}$ and for $ p = 1$ it is called the trace norm or nuclear norm.}\\

{\bf Remark  2.4.}
 {\it a) For $ 1 \leq p \leq 2$, we have the following inegality\\

$\mid\mid K \mid\mid_{p}^{p} \leq \displaystyle{\sum_{n=1}^{\infty} \mid\mid Ke_{n}\mid\mid^{p}.}$ $\hfill { }  (2.1)$\\

Here ${e_{n}}$ is an arbitrary orthogonal basis of the space on which $K$ acting.\\

b) The sum of the series $\displaystyle{\sum_{n=1}^{\infty} < Ke_{n},e_{n} >}$ is said to be matrix trace and denoted by TrK and the sum of the series $\displaystyle{\sum_{n=1}^{\infty} \lambda_{n}}$ is said to be spectral trace of the operator $K$.

c) The operators of class ${\bf C_{1}}$ are called operators with trace because following Lidskii theorem {\bf[18]} these and only these operators have a finite sum of diagonal elements of matrix representation in a certain orthonormal basis and coincides with the spectral trace.}\\

$\displaystyle{\sum_{n=1}^{\infty} < Ke_{n}, e_{n} >}=\displaystyle{\sum_{n=1}^{\infty} \lambda_{n}}.$ $\hfill { }  (2.2)$ \\

{\bf Proposition 2.5.}
{\it 1) The resolvent of the operator $G$ belongs to the class Carlemann ${\bf C_{p}}$,  $\forall \quad p > \frac{1}{3}$.
In particulary the operator $G$ belongs to class of operators with trace resolvent.\\

2) Let $\lambda_{n} = n(n-1)(n-2)$ the eigenvalues of the operator $G$ then \\
a) There exists an infinite sequence of positive integers $n_{m}$ such that:\\

 $\lambda_{n_{m+1}} - \lambda_{n_{m}} \geq c n_{m}^{2}$, $c$ is independent of $n_{m}.$ $\hfill { }  (2.3)$ \\

b) If $\mid n - n_{m}\mid \geq \epsilon n_{m}$, where $\epsilon$ is a fixed, arbitrarily small, positive number, then\\ $\mid\lambda_{n} - \lambda_{n_{m}}\mid \geq c_{\epsilon} inf\{n_{m}^{3},n^{3}\}  > 0$ $\hfill { }  (2.4)$ \\

3) Let $ \lambda_{n} =n(n-1)(n-2)$ for $ n = 3,4,...$ and $\sigma_{m} = \frac{\lambda_{m} + \lambda_{m+1}}{2}$ \\ Then $\{\sigma_{m}\}$ is an increasing sequence of positive numbers $\{\sigma_{m}\}$ such that $\sigma_{m} \rightarrow \infty$ as $ m \rightarrow \infty$ and we have\\

$\displaystyle{\sum_{n=3}^{\infty}\frac{1}{\lambda_{n} - \sigma_{m}} \leq C}$ where $C$ is a positive number.$\hfill { }  (2.5)$ \\\\

4) On the circles $C(0, r_{n_{m}}) = \{ \sigma; \sigma =  r_{n_{m}} = \frac{\lambda_{n_{m}} + \lambda_{n_{m+1}}}{2}\}$ the following upper bound for the trace norm of the resolvent of the operator $G$ is valid:\\

$\displaystyle{\mid\mid(G - \sigma I)^{-1}\mid\mid_{1} \leq \frac{\tilde{c}_{\epsilon}}{n_{m}}}$ $\hfill { }  (2.6)$ \\

 where $\tilde{c}_{\epsilon}$ is independent of $n_{m}$.\\

{\bf Proof}\\

1) As $\lambda_{n} = n(n-1)(n-2)$ then there exist $c_{1}\in ]0, 1[$ and $c_{2} > 0$ such that for large $n$ we have $ c_{1}n^{3} \leq \lambda_{n} \leq c_{1}n^{3}$, in particulary $\frac{1}{\lambda_{n}^{p}} \leq \frac{c_{1}^{p}}{n^{3p}}$, it follows that the series of term general $\frac{1}{\lambda_{n}^{p}}$ converges for all $p > \frac{1}{3}$, i.e. the resolvent of the operator $G$ belongs to Carleman's class  ${\bf C_{p}}$,  $\forall p > \frac{1}{3}$.\\

In particulary the operator $G$ belongs to class of operators with trace resolvent.\\ Note that this elementary proof is valid if $\lambda_{n}\sim n^{\alpha}$ with $\alpha > 0$.\\

2) a) If we suppose the opposite of result of the property a), i.e. for any $\epsilon > 0$ there exists a number $n_{\epsilon}$ such that for all $ n \geq n_{\epsilon}$ we have $\lambda_{n+1} - \lambda_{n} \leq \epsilon n^{2}$ then for all $\epsilon >0$ and $k$, we deduce that\\

$\lambda_{n_{\epsilon}+k} - \lambda_{n_{\epsilon}} \leq \epsilon (n_{\epsilon} + k)^{2} < \epsilon k(n_{\epsilon} + k)^{2} < \epsilon (n_{\epsilon} + k)^{3}$ $\hfill {(2.7) } $\\

Dividing both sides of obtained inequality by $(n_{\epsilon} + k)^{3}$ we get :\\

$\frac{\lambda_{n_{\epsilon}+k}}{(n_{\epsilon} + k)^{3}} - \frac{\lambda_{n_{\epsilon}}}{(n_{\epsilon} + k)^{3}} \leq \epsilon $\\

In particulary\\

$\frac{c_{1}(n_{\epsilon} + k)^{3}}{(n_{\epsilon} + k)^{3}} - \frac{c_{2}(n_{\epsilon})^{3}}{(n_{\epsilon} + k)^{3}} \leq \epsilon$\\

Hence, passing to the limit under $ k \rightarrow \infty$ we obtain $ c_{1} \leq \epsilon$, which contradicts the arbitrariness of $\epsilon$. Thus, $\lambda_{n_{m+1}} - \lambda_{n_{m}} \geq c n_{m}^{2}$ holds for some $c > 0$  and some sequence of numbers $n_{m}$ where $c$ is independent of $n_{m}$.\\

b) If $\mid n - n_{m}\mid\geq \epsilon n_{m}$, i.e. $ n - n_{m}\geq \epsilon n_{m}$ or $ n - n_{m}\leq - \epsilon n_{m}$, we deduce that\\
If $n \geq n_{m}$, we have  $ n^{3} - n_{m}^{3}\geq C_{\epsilon} n_{m}^{3} \geq C_{\epsilon} inf\{n^{3},n_{m}^{3}\}$ where $C_{\epsilon} = \epsilon(\epsilon^{3} + 3\epsilon + 3)$.
\\Hence\\

$ \lambda_{n} - \lambda_{n_{m}} \geq C_{\epsilon} inf\{n^{3},n_{m}^{3}\}.$ $\hfill {(2.8) } $\\

If $n \leq n_{m}$, we have  $ n^{3} - n_{m}^{3}\leq -\tilde{C}_{\epsilon} n_{m}^{3} \leq -\tilde{C}_{\epsilon}inf\{n^{3},n_{m}^{3}\}$\\ where $C_{\epsilon} = \epsilon(\epsilon^{3} - 3\epsilon + 3)$.
\\ Hence \\

$ \lambda_{n} - \lambda_{n_{m}} \leq -C_{\epsilon} inf\{n^{3},n_{m}^{3}\}$.$\hfill {(2.9) } $\\

Now, we put $C= inf\{C_{\epsilon},\tilde{C}_{\epsilon}\}$ to get\\

$\mid \lambda_{n} - \lambda_{n_{m}}\mid \geq C \quad inf\{n^{3},n_{m}^{3}\}.$ $\hfill {(2.10)} $\\

3) Let $ \lambda_{n} =n(n-1)(n-2)$ for $ n = 3,4,...$ and $\sigma_{m} = \frac{\lambda_{m} + \lambda_{m+1}}{2}$. Then \\

$\displaystyle{\sum_{n=3}^{\infty}\frac{1}{\lambda_{n} - \sigma_{m}}} = \displaystyle{\sum_{n=3}^{m}\frac{1}{ \sigma_{m} - \lambda_{n} - \sigma_{m}}} + \displaystyle{\sum_{n= m+1}^{\infty}\frac{1}{\lambda_{n} - \sigma_{m}}}$ $\hfill { }(2.11) $\\

Let us estimate $\displaystyle{\sum_{n= m+1}^{\infty}\frac{1}{\lambda_{n} - \sigma_{m}}}$, if we put $ n-m = k$, we obtain\\

$\displaystyle{\sum_{n= m+1}^{\infty}\frac{1}{\lambda_{n} - \sigma_{m}}} = \displaystyle{\sum_{k= 1}^{\infty}\frac{1}{\lambda_{m+k} - \sigma_{m}} }= \displaystyle{\sum_{k= 1}^{\infty}\frac{2}{\lambda_{m+k} - \lambda_{m} + \lambda_{m+k} - \lambda_{m+1}}}$$\hfill { }(2.12) $\\

Let us estimate $\lambda_{m+k} - \lambda_{m}$ for sufficiently large $m$.\\

Consider the sequence\\

$f(m) = \lambda_{m+k} - \lambda_{m} = (m+k)(m+k-1)(m+k-2) - m(m-1)(m-2)$ for $ m\geq 3$.\\
Then we have\\
$f'(m) = (m+k-1)(m+k-2) + (m+k)(m+k-2) + (m+k)(m+k-1) -(m-1)(m-2)- m(m-2) - m(m-1)$\\
and\\
$f''(m)  = 6k > 0 $ for $k \geq 1$ \\

It follows that $f'(m)$ increases and as $f'(3) = 3k^{2} + 11k -2$ takes positive values then it also follows that $f(m)$ increases.\\

As $f(3) = (k+3)(k+2)(k+1) - 6 = k^{3} + 6k^{2} + 11k$ then there exists $m_{0} = 3$ such that for $m > m_{0}$ we have \\

$\lambda_{m+k} - \lambda_{m} \geq \lambda_{m_{0}+k} - \lambda_{m_{0}}\sim k^{3}$ for $k \rightarrow \infty$.\\

Therefore for sufficiently large $m$, we have\\

$\displaystyle{\sum_{n= 3}^{\infty}\frac{1}{\lambda_{n} - \sigma_{m}} \leq } \displaystyle{\sum_{k= 1}^{\infty}\frac{2}{\lambda_{m+k} - \lambda_{m}} + } \displaystyle{\sum_{k= 2}^{\infty}\frac{2}{\lambda_{m+k} - \lambda_{m+1}} \leq M}\displaystyle{\sum_{k= 1}^{\infty}\frac{1}{k^{3}} = C}$ $\hfill { }(2.13) $\\

($C$ does not depend on $m$).\\

Now let us estimate $\displaystyle{\sum_{n= 3}^{m}\frac{1}{\sigma_{m} - \lambda_{n}}}$.We put $n-3 =k$ to get\\

$$\displaystyle{\sum_{n= m+1}^{\infty}\frac{1}{\sigma_{m} -\lambda_{n}} =} \displaystyle{\sum_{k= 0}^{m-3}\frac{1}{\sigma_{m} - \lambda_{m-k} } = } \displaystyle{\sum_{k=0}^{m-3}\frac{2}{ \lambda_{m} -\lambda_{m-k} + \lambda_{m+1} - \lambda_{m-k}}}$$\\

$$\quad \quad \quad\leq \displaystyle{\sum_{k=1}^{m-3}\frac{2}{ \lambda_{m} -\lambda_{m-k} } + }\displaystyle{\sum_{k=0}^{m-3}\frac{2}{\lambda_{m+1} - \lambda_{m-k}}}$$\\

Consider the term $\lambda_{m} -\lambda_{m-k} =\lambda_{m-k+k} -\lambda_{m-k}$ for $ m-k \geq 3$, then we have\\

$\lambda_{m} -\lambda_{m-k} \geq\lambda_{3+k} -\lambda_{3}\sim k^{3}$. \\

Hence\\

$\displaystyle{\sum_{k=1}^{m-3}\frac{2}{ \lambda_{m} -\lambda_{m-k} }\leq C } \displaystyle{\sum_{k=1}^{\infty}\frac{1}{k^{3}}}$ $\hfill {(2.14) } $\\

where $C$ is a constant does not depend on $m$.\\

 Similarly, the sum $\displaystyle{\sum_{k=0}^{m-3}\frac{2}{ \lambda_{m+1} -\lambda_{m-k}}}$ is bounded.\\ This completes the proof of existence of sequence of positive numbers $\{\sigma_{m}\}$ which increases $\sigma_{m} \rightarrow \infty$ as
 $m \rightarrow \infty$ such that\\

$\displaystyle{\sum_{n=3}^{\infty}\frac{1}{\lambda_{n} - \sigma_{m}} \leq C}$ $\hfill {(2.15) } $\\

where $C$ is a positive number.\\

4) For $\sigma \in C(0, r_{n_{m}})$, we have\\

$\mid\mid(G - \sigma I)^{-1}\mid\mid_{1} = \displaystyle{\sum_{n=1}^{\infty}\frac{1}{\mid \sigma - \lambda_{n}\mid}} = \displaystyle{\sum_{n=1}^{n_{m} -[\epsilon n_{m}]}\frac{1}{\mid \sigma - \lambda_{n}\mid} +} \displaystyle{\sum_{n=n_{m} -[\epsilon n_{m}] + 1}^{n_{m} +[\epsilon n_{m}]}\frac{1}{\mid \sigma - \lambda_{n}\mid} + }\displaystyle{\sum_{n=n_{m} +[\epsilon n_{m}] + 1}^{\infty} \frac{1}{\mid \sigma - \lambda_{n}\mid}}$\\

 = ${\bf I_{1}} + {\bf I_{2}} + {\bf I_{3}}$\\

 where $[\epsilon n_{m}]$ represents the greatest integer less than or equal to $\epsilon n_{m}$.\\

 With the help of property 2) of this lemma, by a) we deduce that \\

 $ {\bf I_{2}} \leq \frac{C_{2}}{n_{m}}$ $\hfill {(2.16) } $\\

 and by the property b), we find that\\

 $ {\bf I_{1}} \leq \frac{C_{1}}{n_{m}^{2}}$ $\hfill {(2.17) } $\\

 and\\

 $ {\bf I_{3}} \leq \frac{C_{3}}{n_{m}^{2}}$$\hfill {(2.18) } $\\

 where $C_{i}, i=1,2,3$ are independent of $n_{m}$.$\hfill { }  \Box $\\

Before giving  a generalization of the property 3) of the above proposition to sequence $\lambda_{n} = n^{\beta}(n-1)^{\beta}(n-2)^{\beta}$ where $\beta$ will be the degree of subordination of the perturbation operator $H_{\mu,\lambda}$ to the unperturbed operator $G$}, we have need to recall the notion of subordination of an perturbation operator $B$ to the unperturbed operator $L$ where $L$ and $ B $ are linear operators acting on a Hilbert space $E$ and some usual spectral properties of
$H = H_{\lambda{''},0,\mu,\lambda}$ under lemmas form\\

{\bf Definition  2.6.}
{\it Let $L$ and $B$ be linear operators in a Hilbert space $E$.\\
We say that $B$ is $p$-subordinate $(p \in [0, 1])$ to $L$ if $D(L)\subset D(B)$ and
there exists a strictly positive constant $C$ such that
$\mid\mid B\phi \mid\mid \leq$ C $\mid\mid L\phi \mid\mid^{p}\mid\mid \phi \mid\mid^{1-p}$ for every $\phi \in D(L)$\\
For $p = 1$, we say that $B$ is subordinate to $L$.}\\

{\bf Definition  2.7.}
{\it Let $L$ be a linear operator in a Hilbert space $E$. The
operator $B$ is said to be $L$-compact with order $p \in [0, 1]$ if $D(L)\subset D(B)$
and for any for all $\epsilon > 0 $, there exists a constant $C_{\epsilon} > 0$ such that
$\mid\mid B\phi \mid\mid \leq$  $\epsilon \mid\mid L\phi \mid\mid^{p}\mid\mid \phi \mid\mid^{1-p} + C_{\epsilon} \mid\mid\phi \mid\mid$ for every $\phi \in D(L)$\\
The operator $B$ is called $L$-compact, if $B$ is $L$-compact with unit order.}\\

From these definitions, we deduce the following results:\\

1) Let $L$ be an operator in $E$ with a dense domain $D(L)$ and at
least one regular point $\sigma$.\\

We suppose that:\\

$(\alpha)$  $D(L)\subset D(B)$.\\

$(\beta)$  For any $\epsilon > 0 $, there exists a constant $C_{\epsilon} > 0$ such that\\
$\mid\mid B\phi \mid\mid \leq$  $\epsilon \mid\mid L\phi \mid\mid^{p}\mid\mid \phi \mid\mid^{1-p} + C_{\epsilon} \mid\mid\phi \mid\mid$ for every $\phi \in D(L)$.\\

$(\gamma)$ $(L - \sigma I)^{-1}$ is compact.\\

Then $B(L - \sigma I)^{-1}$ is a compact operator.\\

2) Let $L$ be an operator in $E$ with a dense domain $D(L)$ and at
least one regular point $\sigma$.\\

We suppose that:\\

$(\alpha)$  $D(L)\subset D(B)$.\\

$(\beta)$ $B(L - \sigma I)^{-1}$ is compact.\\

Then for any for all $\epsilon > 0 $, there exists a constant $C_{\epsilon} > 0$ such that \\

$\mid\mid B\phi \mid\mid \leq$  $\epsilon \mid\mid L\phi \mid\mid^{p}\mid\mid \phi \mid\mid^{1-p} + C_{\epsilon} \mid\mid\phi \mid\mid$ for every $\phi \in D(L)$.\\

Now we recall some usual spectral properties of $ H = H_{\lambda{''},0,\mu,\lambda}$ under lemma form\\

{\bf Lemma 2.8.}
{\it Let $ G = a^{*3}a^{3}, H_{\mu,\lambda} = \mu a^{*}a + i\lambda a^{*}( a + a^{*})a, \tilde{G} = G + I$ and \\
$ H = \lambda^{''}\tilde{G} + H_{\mu,\lambda}$ then\\

1) $Re< H\phi, \phi > = \lambda{''}\mid\mid a^{3}\phi \mid\mid^{2} + \lambda{''}\mid\mid \phi \mid\mid^{2} + \mu\mid\mid a\phi \mid\mid, \forall \phi \in D(H)$.\\

 In particular, for $\lambda^{''} > 0$ we have\\

$(\lambda^{''} + \mu)\mid\mid \phi \mid\mid \leq \mid\mid H\phi \mid\mid , \forall \phi \in D(G)$.\\

2) For $\mu > 0$, $H_{\mu,\lambda}$ is $\frac{1}{2}$-subordinate to $\tilde{G}$ (see lemma 3.1 [1]) i.e. \\
$\exists C_{\mu,\lambda} > 0; \mid\mid H_{\mu,\lambda}\phi\mid\mid \leq C_{\mu,\lambda}\mid\mid \tilde{G}\phi \mid\mid^{\frac{1}{2}} \mid\mid \phi \mid\mid^{\frac{1}{2}} \forall \phi \in D(G).$\\

 In particular \\

 $\exists c_{1} > 0, c_{2} > 0 ; \mid\mid H_{\mu,\lambda}\phi\mid\mid \leq c_{1}\mid\mid \tilde{G}\phi \mid\mid + c_{2}\mid\mid \phi\mid\mid \forall \phi \in D(G)$.\\

3) If $\mu > 0$, the eigenvalues of the operator $H_{\mu,\lambda} $ are reals.\\

4) Let $ 3\leq \beta < 4$, then for each $\epsilon > 0 $, there exists a constant $C_{\epsilon} > 0$ such that \\

$\mid\mid H_{\mu,\lambda}\phi \mid\mid \leq$  $\epsilon \mid\mid (G+I)\phi \mid\mid^{\frac{2}{\beta}}\mid\mid \phi \mid\mid^{1-\frac{2}{\beta}} + C_{\epsilon} \mid\mid\phi \mid\mid$ $\forall$ $\phi \in D(G)$ $\hfill { }  (2.19)$\\

{\bf Proof }\\

1) In Bargmann representation, although the operator $a^{*}( a + a^{*})a$ is non-self-adjoint, it easy to see it is symmetric that implies that\\

 $Re< H\phi, \phi > = \lambda{''}\mid\mid a^{3}\phi \mid\mid^{2} + \lambda{''}\mid\mid \phi \mid\mid^{2} + \mu\mid\mid a\phi \mid\mid, \forall \phi \in D(H)$\\

 By using the well known inequality \\

$ \mid \mid \phi \mid\mid \leq \mid \mid a\phi \mid\mid \forall \phi \in D(a) \bigcap E_{0}$\\

where  $E_{0} = \{ \phi \in E; \phi(0) = 0\}$\\

and the Cauchy-inequality to get\\

$ (\lambda^{''} + \mu)\mid\mid \phi \mid\mid \leq \mid\mid H\phi \mid\mid \forall \phi \in D(G)$\\

2) It be shown in {\bf [1]} (see lemma 3.1).\\

3) Although the operator $H_{\mu,\lambda} $ is non-self-adjoint, it is shown in {\bf[10]}  that its spectrum is real.\\

4) This property is shown in {\bf[1]} (see lemma 3.4).$\hfill { }  \Box $\\

{\bf Lemma 2.9.}
{\it  The operator $ H_{\mu,\lambda}( G - \sigma I)^{-1}$ is nuclear on Bargmann space where $\sigma$ belongs a resolvent set of the operator $G$.}\\

{\bf Proof}\\

 Consider $\phi(z) = \displaystyle{\sum_{n=1}^{\infty}\phi_{n}e_{n}(z)}$ in Bargmann space $E$, The matrix of the operator $H_{\mu,\lambda}$ in basis  ${e_{n}(z)}$ has the form \\

$(H_{\mu,\lambda}\phi)_{n} = \alpha_{n-1}\phi_{n-1} + q_{n}\phi_{n} +\alpha_{n}\phi_{n+1}, n\geq 2 $ $\hfill {(2.20) } $\\

with the initial condition\\

$(H\phi)_{1} = q_{1}\phi_{1} + \alpha_{1}\phi_{2}$ $\hfill {(2.21) } $\\

where\\

$q_{n} = \mu n$, and $\alpha_{n} = i\lambda n\sqrt{n+1}$, ( $\mu$ and $\lambda$ are real numbers and $i^{2} = -1$).\\

It is complex symmetric tri-diagonale matrix (but not Hermitian!) of the form \\

-$\cases{H_{\mu,\lambda} = (h_{m,n})_{m,n=1}^{\infty}\quad with \quad the \quad elements\cr h_{nn} = \mu n\cr h_{n,n+1}=h_{n+1,n}= i\lambda n\sqrt{n+1}; n = 1,2,...\cr \quad and \cr h_{mn} = 0 \quad for \quad \mid m - n \mid > 1 \cr}$ $\hfill { }  (2.22)$\\

Then\\

 All elements of the matrix $H_{\mu,\lambda}$ have order $O(n^{\frac{3}{2}})$ as $n \rightarrow \infty$ and as all elements of the matrix $(G -\sigma I)^{-1}$ have order $O(n^{-3})$ as $n \rightarrow \infty$ then the elements of the matrix $ H_{\mu,\lambda}(G -\sigma I)^{-1}$ have order $O(n^{\frac{-3}{2}}) $ as $ n \rightarrow \infty$ therefore $ H_{\mu,\lambda}(G -\sigma I)^{-1}$ is a nuclear operator.$\hfill { }  \Box $\\

In the following, we look for $\delta \in I\!\!R$ such that the operators $ H_{\mu,\lambda}\tilde{G}^{-\delta}$ and $ \tilde{G}^{\delta}(\tilde{G} - \sigma I)^{-1}$ are bounded and one of them is nuclear. This choice prevents us from working with operators belonging to some Carlemann (Schatten–von Neumann) class of finite order $p > 1$.\\

 For this objective, we establish the following lemma \\

{\bf Corollary 2.10.}
{\it 1) Let be $\delta \in I\!\!R$ such that $ \frac{1}{2} \leq \delta < \frac{2}{3}$ then the operator $H_{\mu,\lambda}\tilde{G}^{-\delta}$ is bounded and the operator $\tilde{G}^{\delta}(\tilde{G} - \sigma I)^{-1}$ is nuclear ; $\sigma  \in \rho (\tilde{G})$.\\
2) Let be $\delta \in I\!\!R$ such that $ \frac{5}{6} < \delta \leq 1$, the operator $H_{\mu,\lambda}\tilde{G}^{-\delta}$ is nuclear and the operator $\tilde{G}^{\delta}(\tilde{G} - \sigma I)^{-1}$ is bounded ; $\sigma  \in \rho (\tilde{G})$.}\\

{\bf Proof }\\

The matrix of the operator $H_{\mu,\lambda}\tilde{G}^{-\delta}$ in the base $e_{n}(z)$ is tridioagonal and its elements have order $O(n^{\frac{3}{2}-3\delta})$ as $ n \rightarrow \infty $ then it easy to see that,if $ \frac{1}{2} \leq \delta $ then we obtain the property 1) and if $ \frac{5}{6} < \delta \leq 1$, the operator $H_{\mu,\lambda}\tilde{G}^{-\delta}$ is nuclear and the operator $\tilde{G}^{\delta}(\tilde{G} - \sigma I)^{-1}$ is bounded \\ where $\sigma $ belongs to $\rho(\tilde{G})$.$\hfill { }  \Box $\\

In the following, we estimate $\mid \mid H_{\mu,\lambda}(\tilde{G} - \sigma)^{-1} \mid\mid_{1}$. We first give the following lemma we will use.\\

{\bf Lemma 2.11.}
{\it i) The inequality $$\mid \frac{a^{\delta}b^{\epsilon}(a^{1-\delta -\epsilon} - b^{1-\delta -\epsilon})}{a - b}\mid \leq 1 $$ holds for any different numbers $ a , b > 0$ and any $\epsilon, \delta \in [0, 1]$\\

ii) Let $\lambda_{n} = n(n-1)(n-2)$ for $ n = 3, 4, ....$ and $ \beta > \frac{1}{3}$ . Then there exists an increasing sequence of positive numbers $\{\gamma_{m}\}$ such that $\gamma_{m} \rightarrow \infty $ as $ m \rightarrow \infty $ and\\
 $$ \displaystyle{\sum_{n=3}^{+\infty}\frac{1}{\mid \lambda_{n}^{\beta} - \gamma_{m}\mid} \leq C}$$\\
 where $C$ is a positive number.}\\

{\bf Proof }\\

i) This inequality is classically well-known. Note that for $\epsilon = 0$ we have\\

$\mid \frac{a^{\delta}(a^{1-\delta} - b^{1-\delta})}{a - b}\mid \leq 1 $ $\hfill { }  (2.23)$\\
holds for any different numbers $ a , b > 0$ and any $\delta \in [0, 1]$.\\

ii) For $\beta = 1$ this property is the property 3) of proposition 2.5.\\

Now let $m \geq 3$ and $\gamma_{m} = \frac{\lambda_{m}^{\beta} + \lambda_{m+1}^{\beta}}{2}$ with $\lambda_{m}^{\beta} = m^{\beta}(m-1)^{\beta}(m-2)^{\beta}$ and $\beta > \frac{1}{3}$\\
Then \\
$ \displaystyle{\sum_{n=3}^{+\infty}\frac{1}{\mid \lambda_{n}^{\beta} - \gamma_{m}\mid}}$ = $ \displaystyle{\sum_{n=3}^{m}\frac{1}{ - \gamma_{m} - \lambda_{n}^{\beta}}}$ + $ \displaystyle{\sum_{n=m+1}^{+\infty}\frac{1}{\lambda_{n}^{\beta} - \gamma_{m}}}$\\

Let $ n-m = k$ then \\

$\displaystyle{\sum_{n=m+1}^{+\infty}\frac{1}{\lambda_{n}^{\beta} - \gamma_{m}}}$ = $\displaystyle{\sum_{k =1}^{+\infty}\frac{1}{\lambda_{m + k}^{\beta} - \gamma_{m}}}$ = $\displaystyle{\sum_{k =1}^{+\infty}\frac{2}{2\lambda_{m + k}^{\beta} - \lambda_{m}^{\beta} - \lambda_{m + 1}^{\beta}}}$\\

Now to estimate $\lambda_{m + k}^{\beta} - \lambda_{m}^{\beta}$ we consider the function \\

$f_{\beta}(x) = x^{\beta}(x-1)^{\beta}(x-2)^{\beta}$ for $x \geq 3$ $\hfill {(2.24) } $\\

It follows that\\

$f^{'}_{\beta}(x) = \beta f_{\beta - 1}(x)(3x^{2} - 3x + 2)$ \\

and\\

$f^{''}_{\beta}(x) = f_{\beta - 2}(x)[ (9\beta (\beta -1) + 6\beta)x^{4} - (18\beta (\beta -1) - 21\beta) x^{3} + 21\beta^{2}x^{2} - (12\beta (\beta -1) + 6\beta)x + 4\beta (\beta -1)]$\\

For $9\beta (\beta -1) + 6\beta \geq 0$ in particulary for $ \beta > \frac{1}{3}$ and sufficiently large $x$ , it follows that  $f^{''}_{\beta}(x)$ take positive values for sufficiently large $x$ ($ x > x_{min} \geq 3$) then it follows that $f^{'}_{\beta}(x)$ increases and in particulary $f^{'}_{\beta}(x) < f^{'}_{\beta}(x + h)$ for all $ h > 0$  and for $x > x_{min}$ therefore $ f_{\beta}(x_{min} + h) - f_{\beta}(x_{min}) \leq f_{\beta}(x + h) - f_{\beta}(x)$ for $ x \geq x_{min}$.\\
Thus there exists a number $m_{0}$ depending only on $\beta$ such that, for $ m > m_{0} \geq 3$ we have :\\

$\lambda_{m + k}^{\beta} - \lambda_{m}^{\beta} \geq \lambda_{m_{0} + k}^{\beta} - \lambda_{m_{0}}^{\beta}$\\

By noting that $\lambda_{m_{0} + k}^{\beta} - \lambda_{m_{0}}^{\beta}\sim k^{3\beta}$ and \\

$\displaystyle{\sum_{k =1}^{+\infty}\frac{2}{2\lambda_{m + k}^{\beta} - \lambda_{m}^{\beta} - \lambda_{m + 1}^{\beta}}}$ = $\displaystyle{\sum_{k =1}^{+\infty}\frac{2}{\lambda_{m + k}^{\beta} - \lambda_{m}^{\beta} + \lambda_{m + k}^{\beta} - \lambda_{m + 1}^{\beta}}}$
= $\displaystyle{\frac{2}{\lambda_{m + 1}^{\beta} - \lambda_{m}^{\beta}}}$ + $\displaystyle{\sum_{k =2}^{+\infty}\frac{2}{\lambda_{m + k}^{\beta} - \lambda_{m}^{\beta} + \lambda_{m + k}^{\beta} - \lambda_{m + 1}^{\beta}}}$

$\displaystyle{\quad \quad \quad \quad \quad \leq \frac{2}{\lambda_{m + 1}^{\beta} - \lambda_{m}^{\beta}}}$ + $\displaystyle{\sum_{k =2}^{+\infty}\frac{2}{\lambda_{m + k}^{\beta} - \lambda_{m}^{\beta}}}$ + $\displaystyle{\sum_{k=2}^{+\infty}\frac{2}{\lambda_{m + k}^{\beta} - \lambda_{m + 1}^{\beta}}}$ \\

$\quad \quad \quad \quad \quad \leq $ $\displaystyle{\sum_{k =1}^{+\infty}\frac{2}{\lambda_{m + k}^{\beta} - \lambda_{m}^{\beta}}}$ + $\displaystyle{\sum_{k =2}^{+\infty}\frac{2}{\lambda_{m + k}^{\beta} - \lambda_{m + 1}^{\beta}}}$ \\

$\quad \quad \quad \quad \quad \quad \quad \leq C_{1}$ $\displaystyle{\sum_{k =1}^{+\infty}\frac{1}{k^{3\beta}} = C}$ ($C_{1}$ does not depend on $m$).$\hfill { }  \Box $\\

we deduce that \\

{\bf Theorem 2.12.} {\it (estimatation of $\mid \mid H_{\mu,\lambda}(\tilde{G} - \sigma)^{-1} \mid\mid_{1}$)\\
For $ \frac{1}{2} \leq \delta < \frac{2}{3}$, then for all $\alpha$ ; $ 0 \leq \alpha < \frac{2}{3} - \delta $ there exists a sequence of numbers $\eta_{m}$ such that $\lambda_{m} < \eta_{m} < \lambda_{m+1}$ and for $\mid \sigma \mid = \eta_{m}$ we have $\mid \mid H_{\mu,\lambda}(G - \sigma I)^{-1}\mid\mid_{1} = o(\frac{1}{\eta_{m}^{\alpha}})$  where  $\mid\mid . \mid\mid_{1}$ is the nuclear norm.}\\

{\bf Proof }\\

To estimate $\mid \mid H_{\mu,\lambda}(G - \sigma I)^{-1}\mid\mid_{1}$, we note that\\
$\mid \mid H_{\mu,\lambda}(G - \sigma I)^{-1}\mid\mid_{1} \leq \mid \mid H_{\mu,\lambda}G^{-\delta}\mid\mid.\mid \mid G^{\delta}(G - \sigma I)^{-1}\mid\mid_{1} = \mid \mid H_{\mu,\lambda}G^{-\delta}\mid\mid \displaystyle{\sum_{n=1}^{\infty}\frac{\lambda_{n}^{\delta}}{\mid \lambda_{n} - \sigma\mid}}$\\
 $\leq\mid \mid H_{\mu,\lambda}G^{-\delta}\mid\mid \displaystyle{\sum_{n=1}^{\infty}\frac{\lambda_{n}^{\delta}}{\mid \lambda_{n} - \mid\sigma\mid\mid}}$\\

 By using the inequality $$\mid \frac{a^{\delta}b^{\alpha}(a^{1-\delta -\alpha} - b^{1-\delta -\alpha})}{a - b}\mid \leq 1 $$ we get\\

 $\mid \mid H_{\mu,\lambda}(G - \sigma I)^{-1}\mid\mid_{1} \leq \mid \mid H_{\mu,\lambda}G^{-\delta}\mid\mid $ $\displaystyle{\sum_{n=1}^{\infty}\frac{\lambda_{n}^{\delta}}{ \lambda_{n}^{\delta}\mid \sigma\mid^{\alpha}\mid \lambda_{n}^{1-\delta-\alpha} - \mid\sigma\mid^{1-\delta-\alpha}\mid}}$\\

 $$=\frac{\mid \mid H_{\mu,\lambda}G^{-\delta}\mid\mid }{\mid\sigma\mid^{\alpha}} \displaystyle{\sum_{n=1}^{\infty}\frac{1}{\mid \lambda_{n}^{1-\delta-\alpha} - \mid\sigma\mid^{1-\delta-\alpha}\mid}}$$\\

 From the relation $\lambda_{n}^{1-\delta-\alpha}\sim n^{3(1-\delta-\alpha)}$ then for $0 \leq \alpha < \frac{2}{3}- \delta$ the operator $G^{-(1-\delta-\alpha)}$  is nuclear and the series converges.\\

 By choosing $\eta_{m} = \displaystyle{[\frac{\lambda_{m}^{1-\delta-\alpha} + \lambda_{m+1}^{1-\delta-\alpha}}{2}]^{\frac{1}{1-\delta-\alpha}}}$ and applying the above lemma we deduce that\\
 $\displaystyle{\sum_{n=1}^{\infty}\frac{1}{\mid \lambda_{n}^{1-\delta-\alpha} - \eta_{m}^{1-\delta-\alpha}\mid}} \leq C$$\hfill {(2.25) } $\\

  where $C$ does not depend on $m$\\

  The arbitrariness of the choice of $\alpha$ allows us to deduce that for $\mid \sigma \mid = \eta_{m}$ we have\\

  $\mid\mid H_{\mu,\lambda}(G - \sigma I)^{-1}\mid\mid_{1}$ =
  $o(\frac{1}{\eta_{m}^{\alpha}}) $ as $ m \rightarrow \infty $ $\hfill {(2.26)} $

\begin{center}
{\bf 3. Explicit calculating of regularized trace of magic Gribov operator}\\
\end{center}

To determine the number of corrections necessary for the existence of finite formula of the regularized trace of our operator, we have need to recall some results on the infinite determinants and some relations of the theory of perturbation determinants.\\

{\bf A) Some Reviews on the infinite determinants and some relations of the theory of perturbation determinants.}\\

Given a trace class operator $K$ acting on a Hilbert space i.e. $K \in C_{1}$, then its trace is given by Lidskii's theorem\\
$$Tr(K) = \displaystyle{\sum_{n=1}^{\infty} \lambda_{n} < \infty}$$ Now the Fredholm determinant of
the operator $ I + K$ is defined correctly by\\

$ det(1 + K) = \displaystyle{\prod_{n=1}^{\infty}(1 + \lambda_{n})} $ $\hfill {(3.1)}$ \\

where $I$ is the identity operator and the numbers $\lambda_{n}$ are the eigenvalues of $K$, repeated
the times indicated by their corresponding multiplicities.\\
There are several equivalent definitions for $det(I + K)$ for $K \in C_{1}$. For example for any $z \in I\!\!\!\!C$ we have \\

$ det(1 + zK) = \displaystyle{\prod_{n=1}^{\infty}(1 + z\lambda_{n})}$ $\hfill {(3.2)}$ \\

$ det(1 + zK) = e^{Tr Log(I + zK)}$ $\hfill {(3.3)}$ \\

The latter definition is only determined modulo $2i\pi$ and it leads to the small $z$ expansion known as Plemelj's formula:\\

$ det(1 + zK) = e^{\displaystyle{\sum_{m=1}^{\infty}(-1)^{m}z^{m}\frac{TrK^{m}}{m}}}$ $\hfill {(3.4)}$ \\

which converges if $Tr\sqrt{K^{*}K} < 1$. The equivalence of these three definitions is established
through Lidskii's theorem\\

There are two important properties of the determinant so defined.\\
First the multiplication formula\\

$ det(I + K_{1} + K_{2} + K_{1}K_{2}) = det(I + K_{1}).det(I + K_{2})$ $\hfill {(3.5)}$\\

holds for all $K_{1};K_{2} \in C_{1}$.\\

Second, the characterization of invertibility:\\

$ det(I + K) \neq 0 $ if and only if $(I + K)^{-1}$ exists.\\

Also we can recall the the integral representation of the Fredholm determinant  which is given by\\

$ det(1 + K) = e^{\int_{\gamma}Tr[K(I + \sigma K)^{-1}]}d\sigma$ $\hfill {(3.6)}$ \\

with\\
$\gamma: [0; 1] \rightarrow I\!\!\!\!C$ a continuous path such that
$\gamma(0) = 0 ,\gamma(1) = 1$ and that the operator $(I + \sigma K)^{-1}$ exists and is bounded for all $\sigma$ in $\gamma$\\

Some properties connected with the differentiability of the Fredholm determinants are
recalled now.\\

Let $\Omega$ an open subset of $I\!\!\!\!C$ and $F(z)$ a holomorphic application over $\Omega$  taking values on the ideal of the trace class operators such that the trace norm $\mid\mid F(z)\mid\mid_{1} $ of $F(z)$ is bounded over each compact subset of $\Omega$ then we recall some properties of  $F(z)$ under lemma form.The corresponding proofs can be found in (Gohberg Krein {\bf[8]}, Ch IV, p.156-171).\\

{\bf Lemma 3.1.} {\it
Under the above hypotheses, we have\\
1) The funtion $det(I + F(z)) : \Omega \rightarrow I\!\!\!\!C $ is holomorphic.\\
2) The derivative of $F(z)$ is a trace class operator for all $z \in \Omega$.\\
3) The funtion $Tr(F(z))$ is holomorphic on $\Omega$.\\
4) $\frac{d}{dz}Tr(F(z)) = Tr[\frac{d}{dz}(F(z))]$.\\
5) $\frac{d}{dz}Log(det(I + F(z)) = Tr[(I + F(z))^{-1}\frac{d}{dz}(F(z))]$.}}\\

{\bf Remark 3.2.}
{\it For a pair of operators $A_{0}$ and $A = A_{0} + B $   where $R(\lambda)$ denotes the resolvent of $ A$ and $R_{0}(\lambda)$ denotes the resolvent of $ A_{0}$ . If $BR_{0}(\lambda)$ is a nuclear operator, it is possible to introduce the so-called perturbation determinant:\\

$D_{A/A_{0}(\lambda)} = det[(A - \lambda I) · (A_{0} - \lambda I)^{-1}] = det[I + BR_{0}(\lambda)].$ $\hfill {(3.7)}$\\

We will need two relations for this determinant (Gohberg-Krein {\bf[8]}, Ch IV, p.171-173):\\

$Tr(R_{0}(\lambda) - R(\lambda)) = \frac{d}{d\lambda}(LnD_{A/A_{0}(\lambda)})$ $ \hfill { }  (3.8)$\\

$LnD_{A/A_{0}}(\lambda) = Tr Ln(I + BR_{0}(\lambda))$ $\hfill { }  (3.9)$\\

Most of the results on infinite determinants of Hilbert space operators can be founded in ({\bf[7]}, {\bf[20]}, {\bf[26]} or {\bf[27]}).}\\

{\bf B) Proof of main theorem}\\

In the following, we will find some relations about the eigenvalues and resolvent of the operators $\lambda{''}G$ and $H = \lambda{''}G + H_{\mu,\lambda}$. These operators have purely-discrete spectrum.\\ Moreover the resolvent $R_{\sigma}^{0} = (\lambda^{''}G -\sigma I)^{-1}$ and $R_{\sigma} = (\lambda^{''}G  + H_{\mu,\lambda} -\sigma I)^{-1}$ are connected by the relation\\

$R_{\sigma} = R_{\sigma}^{0} + R_{\sigma}^{0}\displaystyle{\sum_{k=1}^{\infty }(-1)^{k}[H_{\mu,\lambda}R_{\sigma}^{0}]^{k}}$ $\hfill {(3.10)}$\\

\noindent where the convergence of the series in the right-hand side of above formula is understood in the sense of convergence in the norm of the ring of bounded operators.\\

{\bf Remark 3.3.}
{\it To use the equality (3.9) in the form of a power series for the logarithm at our operator $ H = \lambda{''}G + H_{\mu,\lambda}$, we choose $A_{0} = \lambda^{''}G$ , $ B = H_{\mu,\lambda}$
and $\sigma$ belongs to the circles $\gamma_{m}$ in complex plane of radii $\{\eta_{m}\}$ centered at zero\\
and for $m$ sufficiently large, one hand, it is well known (see theorem 3.3 {\bf[1]}) that $$\mid\mid H_{\mu,\lambda}(\lambda{''}G - \sigma I)^{-1}\mid\mid < 1.$$\\ and the other, following the properties possessed by the operators $\lambda^{''}\tilde{G}$ and $H_{\mu,\lambda}$ allow one to assert that the
family of operators $\lambda^{''}\tilde{G} +\tau H_{\mu,\lambda}$ is a holomorphic family of type (A) {\bf[17}, Ch. VII{\bf]}, and results from the analytical theory of perturbations {\bf[17]} guarantee that eigenvalues of the family of operators $\lambda^{''}\tilde{G} +\tau H_{\mu,\lambda}$ are in any case continuous functions of the parameter $\tau$ .\\
The above argument is also valid for the perturbation $\tau H_{\mu,\lambda}$ for $\tau \in [0, 1]$, hence all $\sigma$ belonging to the circles $\gamma_{m}$ remain in the resolvent set of this family for all $\tau \in [0, 1]$, so eigenvalues of operators $\lambda^{''}\tilde{G} +\tau H_{\mu,\lambda}$ do not intersect the contour $\gamma_{m}$ for $\tau \in [0, 1]$.\\
Then for sufficiently large $m$, inside the contours $\gamma_{m}$ there is an identical number of eigenvalues of the operators $\lambda^{''}G + H_{\mu,\lambda}$ and $\lambda^{''}G$.}\\

We denote the eigenvalues  of $H$ by $\{\sigma_{n}\}$ and we continue to denote  the eigenvalues  of $\lambda{''}G$ by $\{\lambda_{n}\}$\\

Now, let $\mu > 0$ and let $\sigma \in I\!\!\!\!C $ such that $\mid \sigma \mid = \frac{\lambda_{n_{m}+1} + \lambda_{n_{m}}}{2} = r_{m}$, we note that \\

i) For large value of $m$ the inequalities $ \sigma_{n_{m}} < r_{m} < \sigma_{n_{m}+1}$ and $ \lambda_{n_{m}} < r_{m} < \lambda_{n_{m}+1}$ are satisfied.\\

ii) The series $\displaystyle{\sum_{n=3}^{\infty }\frac{\sigma}{\tilde{\sigma}_{n} - \sigma}}$ and $\displaystyle{\sum_{n=3}^{\infty }\frac{\sigma}{\lambda_{n} - \sigma}}$ are uniform convergent on the circle $\mid \sigma \mid = r_{m}$\\

And by Cauchy's integral formula for a disk, we have\\

$\displaystyle{\sigma_{n} = -\frac{1}{2i\pi}\int_{\gamma_{n}}\frac{\sigma}{\sigma_{n} - \sigma}d\sigma}$ $\hfill {(3.11)}$\\
and\\
$\displaystyle{\lambda_{n} = -\frac{1}{2i\pi}\int_{\gamma_{n}}\frac{\sigma}{\lambda_{n} - \sigma}d\sigma}$ $\hfill {(3.12)}$\\

Then\\

$\displaystyle{\sum_{n=1}^{m}(\sigma_{n}-\lambda_{n})} $ =
$\displaystyle{-\frac{1}{2i\pi}\int_{\gamma_{m}}\sigma Tr([\lambda{''}G + }$
$\displaystyle{H_{\mu,\lambda} - \sigma I]^{-1} - [\lambda{''}G - \sigma I]^{-1})d\sigma}$ $\hfill {} (3.13)$\\

Now, we can proceed to proving the aim theorem 1.4 given in introduction.\\
We will investigate the right-hand side of the above equality (3.13) following the reasoning and the techniques used by Sadovnichii and  Podol'skii in  {\bf[25]}.\\
From the above inequality (3.13), integrating by parts and taking into account of the formulas (3.8) (the perturbation determinant) and (3.9), we obtain \\

$\displaystyle{-\frac{1}{2i\pi}\int_{\gamma_{m}}\sigma Tr([\lambda{''}G + H_{\mu,\lambda} - \sigma I]^{-1} - [\lambda{''}G - \sigma I]^{-1})d\sigma }$= $\displaystyle{-\frac{1}{2i\pi}\int_{\gamma_{m}}Tr[Log(I + H_{\mu,\lambda}(\lambda{''}G - \sigma I)^{-1})]d\sigma }$= $\displaystyle{-\frac{1}{2i\pi}\int_{\gamma_{m}}[Tr(H_{\mu,\lambda}R_{\sigma}^{0}) + Tr(\displaystyle{\sum_{k=2}^{\infty }\frac{(-1)^{k-1}}{k}(H_{\mu,\lambda}R_{\sigma}^{0})^{k}})]d\sigma.}$\\

Consider the term with separately. Since the elements of the matrix $H_{\mu,\lambda}R_{\sigma}^{0}$ are $O(n^{\frac{-3}{2}})$ as $n \rightarrow \infty $ we know that $H_{\mu,\lambda}R_{\sigma}^{0}$ is a nuclear operator then its trace can be calculated as the matrix trace in the orthonormal basis $\{e_{n}(z)\}$.\\

Note also that in Bargmann representation, we have\\

$< H_{\mu,\lambda}e_{n}, e_{n}> = n\mu$; $n=1,2,...$ $\hfill {(3.14)}$\\

then \\

$\displaystyle{\frac{1}{2i\pi}\int_{\gamma_{m}}Tr(H_{\mu,\lambda}R_{\sigma}^{0})d\sigma }$ = $\displaystyle{\frac{1}{2i\pi}\int_{\gamma_{m}}\sum_{n=1}^{\infty}\frac{n\mu}{\lambda_{n} -\sigma}d\sigma}$ $\hfill {(3.15)}$\\

Then\\

$\displaystyle{\sum_{n=1}^{m}(\tilde{\sigma}_{n}-\lambda_{n} - n\mu)}$ = $\displaystyle{-\frac{1}{2i\pi}\int_{\gamma_{m}}}$Tr($\displaystyle{\sum_{k=2}^{\infty }\frac{(-1)^{k-1}}{k}(H_{\mu,\lambda}R_{\sigma}^{0})^{k}})d\sigma.$ $\hfill {(3.16)}$\\

Now, let us estimate the terms of this series for $k \geq 2$ by writing it in the following form:\\

$\displaystyle{-\frac{1}{2i\pi}\int_{\gamma_{m}}}$Tr($\displaystyle{\sum_{k=2}^{\infty }\frac{(-1)^{k-1}}{k}(H_{\mu,\lambda}R_{\sigma}^{0})^{k}})d\sigma$ = \\ $\displaystyle{\frac{1}{2i\pi}\int_{\gamma_{m}}}$Tr($\displaystyle{(H_{\mu,\lambda}R_{\sigma}^{0})^{2}})d\sigma$ $\displaystyle{-\frac{1}{2i\pi}\int_{\gamma_{m}}}$Tr($\displaystyle{\sum_{k=3}^{\infty }\frac{(-1)^{k-1}}{k}(H_{\mu,\lambda}R_{\sigma}^{0})^{k}})d\sigma.$ $\hfill {(3.17)}$\\

We begin by given an estimation of second series for $k \geq l$ where $l\geq 3$.\\

Now the following estimates are valid.\\

1) Since the s-numbers of the trace class operator $R_{\sigma}^{0}$ are $\{\frac{1}{\mid \lambda_{n} - \sigma \mid}\}$ then the norm of the operator $ H_{\mu,\lambda}R_{\sigma}^{0}$ is estimated as\\

$ Max_{_{\sigma \in \gamma_{m}}} \mid\mid H_{\mu,\lambda}R_{\sigma}^{0}\mid\mid \leq \mid\mid H_{\mu,\lambda}G^{-\delta}\mid\mid Max_{_{\sigma \in \gamma_{m}}}(Max_{_{n}}\frac{\lambda_{n}^{\delta}}{\mid \lambda_{n} - \sigma \mid})\leq const.Max_{_{n}}\frac{\lambda_{n}^{\delta}}{\mid \lambda_{n} - a_{m} \mid}$.\\

$\leq const.Max_{_{n}}\frac{1}{\mid \lambda_{n}^{1-\delta} - a_{m}^{1-\delta} \mid}$ where $ \lambda_{m} < a_{m} < \lambda_{m+1}$ $\hfill { }  (3.18)$\\

2) Let be $K_{1}$ is bounded operator and $ K_{2} \in C_{1}$ then the estimate $\mid\mid K_{1}K_{2}\mid\mid_{1} \leq  \mid\mid K_{1} \mid\mid .\mid\mid K_{2} \mid\mid_{1}$ is valid.\\

By applying this estimate we get $\mid Tr(H_{\mu,\lambda}R_{\sigma}^{0})^{k}\mid \leq \mid\mid H_{\mu,\lambda}R_{\sigma}^{0}\mid\mid^{k-1} \mid\mid H_{\mu,\lambda}R_{\sigma}^{0}\mid\mid_{1}.$\\
and by last inequality we deduce that \\

$\displaystyle{\int_{\gamma_{m}}\mid Tr(H_{\mu,\lambda}R_{\sigma}^{0})^{k}\mid \mid d\sigma \mid \leq Max_{_{\sigma \in \gamma_{m}}}\mid\mid H_{\mu,\lambda}R_{\sigma}^{0}\mid\mid_{1}\int_{\gamma_{m}}\mid\mid H_{\mu,\lambda}R_{\sigma}^{0}\mid\mid^{k-1} \mid d\sigma \mid}.$ $\hfill { }  (3.19)$\\

3) To estimate the integral $ \displaystyle{\int_{\gamma_{m}}\mid\mid H_{\mu,\lambda}R_{\sigma}^{0}\mid\mid^{k-1} \mid d\sigma \mid}$ we use the techniques of Sadovnichii-Podolskii in {\bf[25]} to get\\

$\displaystyle{\int_{\gamma_{m}}\mid\mid H_{\mu,\lambda}R_{\sigma}^{0}\mid\mid^{k-1} \mid d\sigma \mid \leq \frac{c^{k}\lambda_{m}^{\delta}}{\mid \lambda_{m}^{1-\delta} - a_{m}^{1-\delta}\mid^{k-2}}}$ $\hfill { }  (3.20)$\\

 Then the remainder of the series for $ l \geq 3$ satisfies the relation:\\

 $\mid \displaystyle{-\frac{1}{2i\pi}\int_{\gamma_{m}}Tr(\displaystyle{\sum_{k=l}^{\infty }\frac{(-1)^{k-1}}{k}(H_{\mu,\lambda}R_{\sigma}^{0})^{k}})d\sigma \mid \leq \mid\mid H_{\mu,\lambda}R_{a_{m}}^{0}\mid\mid_{1}}$.$\displaystyle{\int_{\gamma_{m}}\sum_{k=l}^{\infty }\mid\mid H_{\mu,\lambda}R_{\sigma}^{0}\mid\mid^{k-1} \mid\mid \mid d\sigma\mid}$\\

 $\leq \mid\mid H_{\mu,\lambda}R_{a_{m}}^{0}\mid\mid_{1}\lambda_{m}^{\delta} \displaystyle{\sum_{k=l}^{\infty}(\frac{c}{ a_{m}^{1-\delta} - \lambda_{m}^{1-\delta}})^{k-2}}$ \\

 $\leq \displaystyle{C\frac{a_{m}^{\delta}\mid\mid H_{\mu,\lambda}R_{a_{m}}^{0}\mid\mid_{1}}{(a_{m}^{1-\delta} - \lambda_{m}^{1-\delta})^{l-2}}} $ \\

 $\leq \displaystyle{C\frac{a_{m}^{\delta -\alpha(l-2)}\mid\mid H_{\mu,\lambda}R_{a_{m}}^{0}\mid\mid_{1}}{(a_{m}^{1-\delta} - \lambda_{m}^{1-\delta})^{l-2}}}$ = $o\displaystyle{(a_{m}^{\delta - \alpha(l-1)})}$ $\hfill { }  (3.21)$\\

 4) To estimate the integral $\displaystyle{\int_{\gamma_{m}}Tr(H_{\mu,\lambda}R_{\sigma}^{0})^{2})}d\sigma$ it convenient to write the second-order correction in the form\\

 $\displaystyle{\frac{1}{2i\pi}\int_{\gamma_{m}}}$\\

 Tr($\displaystyle{(H_{\mu,\lambda}R_{\sigma}^{0})^{2}})d\sigma$ =
 $\displaystyle{\frac{1}{2i\pi}\int_{\gamma_{m}}}$ 
 $\displaystyle{\sum_{n=1}^{\infty}\frac{<H_{\mu,\lambda}R_{\sigma}^{0}
 H_{\mu,\lambda}e_{n},e_{n}>}{\lambda_{n} - \sigma}d\sigma}$\\

  Let $H_{\mu,\lambda}e_{n} = \displaystyle{\sum_{n=l}^{\infty}<H_{\mu,\lambda}e_{n},e_{k}>e_{k}}$ then by Cauchy theorem we deduce that\\

  $\displaystyle{\frac{1}{2i\pi}\int_{\gamma_{m}}}$Tr($\displaystyle{(H_{\mu,\lambda}R_{\sigma}^{0})^{2}})d\sigma$ = $\displaystyle{\sum_{n=1}^{m}}$ $\displaystyle{\sum_{k=m+1}^{\infty}\frac{<H_{\mu,\lambda}e_{n}, e_{k}><H_{\mu,\lambda}e_{k}, e_{n}>}{\lambda_{k} -\lambda_{n}}}$ $\hfill {(3.22)}$\\

  Now we consider the inner series $\displaystyle{\sum_{k=m+1}^{\infty}\frac{\mid<H_{\mu,\lambda}e_{n}, e_{k}><H_{\mu,\lambda}e_{k}, e_{n}>\mid}{\lambda_{k} -\lambda_{n}}}$ and it convenient to write it in the form\\

 $\displaystyle{\sum_{k=m+1}^{\infty}\frac{\mid<H_{\mu,\lambda}e_{n}, e_{k}><H_{\mu,\lambda}e_{k}, e_{n}>\mid}{\lambda_{k} -\lambda_{n}}}$\\
 =
 $\displaystyle{\sum_{k=m+1}^{\infty}\lambda_{n}^{\delta}\lambda_{k}^{\delta}\frac{\mid<H_{\mu,\lambda}G^{-\delta }e_{n}, e_{k}><H_{\mu,\lambda}G^{-\delta}e_{k}, e_{n}>\mid}{\lambda_{k} -\lambda_{n}}}$\\

 We choose $\epsilon = 0$ in the inequality $\mid \frac{a^{\delta}b^{\epsilon}(a^{1 -\delta -\epsilon} - b^{1-\delta-\epsilon})}{a - b}\mid \leq 1$ to get\\

 $\displaystyle{\frac{\lambda_{k}^{\delta}}{\lambda_{k} - \lambda_{n}} \leq \frac{1}{\lambda_{k}^{1-\delta} - \lambda_{n}^{1-\delta}}}$ $\hfill { } (3.23)$\\\\

 Then we obtain\\

 $\displaystyle{\sum_{k=m+1}^{\infty}\frac{\mid <H_{\mu,\lambda}e_{n}, e_{k}><H_{\mu,\lambda}e_{k}, e_{n}>\mid}{\lambda_{k} -\lambda_{n}}}$\\
 $\leq \lambda_{n}^{\delta}$ $\displaystyle{\sum_{k=m+1}^{\infty}\frac{\mid<H_{\mu,\lambda}G^{-\delta }e_{n}, e_{k}><H_{\mu,\lambda}G^{-\delta}e_{k}, e_{n}>\mid}{\lambda_{k}^{1-\delta} -\lambda_{n}^{1-\delta}}}$ $\hfill {(3.24)}$\\

We will use the Abel transformation to investigate (converging) series by taking \\

$a_{k} = \displaystyle{\sum_{l=m+1}^{k}\mid <H_{\mu,\lambda}e_{n}, e_{l}><H_{\mu,\lambda}e_{l}, e_{n}>}$ this sequence is bounded \\

\noindent and \\

$\displaystyle{ b_{k} = \frac{1}{\lambda_{k}^{1-\delta} -\lambda_{n}^{1-\delta}}}$ this sequence decreases to $0$.\\

Then have \\

$\lambda_{n}^{\delta}\displaystyle{\sum_{k=m+1}^{\infty}\frac{\mid<H_{\mu,\lambda}G^{-\delta }e_{n}, e_{k}><H_{\mu,\lambda}G^{-\delta}e_{k}, e_{n}>\mid}{\lambda_{k}^{1-\delta} -\lambda_{n}^{1-\delta}}}$ =
$\lambda_{n}^{\delta}\displaystyle{\sum_{k=m+1}^{\infty}a_{k}(b_{k}-b_{k+1})}$ \\

Then \\

$\lambda_{n}^{\delta}\displaystyle{\sum_{k=m+1}^{\infty}\frac{\mid<H_{\mu,\lambda}G^{-\delta }e_{n}, e_{k}><H_{\mu,\lambda}G^{-\delta}e_{k}, e_{n}>\mid}{\lambda_{k}^{1-\delta} -\lambda_{n}^{1-\delta}}}$\\

$\leq \lambda_{n}^{\delta}\displaystyle{\sum_{k=m+1}^{\infty}\mid\mid H_{\mu,\lambda}G^{-\delta}\mid\mid^{2}(b_{k}- b_{k+1}) \leq \frac{\lambda_{n}^{\delta}\mid\mid H_{\mu,\lambda}G^{-\delta}\mid\mid^{2}}{b_{m+1}}}$\\

Therefore we have\\

$\displaystyle{\mid \displaystyle{\int_{\gamma_{m}}Tr(H_{\mu,\lambda}R_{\sigma}^{0})^{2}}}d\sigma \mid$ \\

$\leq \displaystyle{\sum_{n=1}^{m}\frac{\lambda_{n}^{\delta}\mid\mid H_{\mu,\lambda}G^{-\delta}\mid\mid^{2}}{b_{m+1}}}$\\

$\leq \lambda_{m}^{\delta - \alpha}\displaystyle{\sum_{n=1}^{m}\frac{\mid\mid H_{\mu,\lambda}G^{-\delta}\mid\mid^{2}}{\lambda_{m+1}^{1 -\delta - \alpha} - \lambda_{n}^{1 -\delta - \alpha}}}$\\

We choose $\alpha$ such that the resolvent of $G^{1-\delta -\alpha}$ is nuclear operator. i.e $G^{-(1-\delta -\alpha)}$ is nuclear then $3(1-\delta -\alpha)> 1$,i.e $ 0\leq \alpha < \frac{2}{3} - \delta $\\

{\bf Remark  3.4}
{\it As $\lambda_{m} < \omega_{m} < \lambda_{m+1}$ Then\\

i) $\lambda_{m}^{\delta - \alpha} < \omega_{m}^{\delta - \alpha} < \lambda_{m+1}^{\delta - \alpha}$\\

ii) $ \omega_{m}^{1-\delta - \alpha} < \lambda_{m+1}^{1-\delta - \alpha}$ and $ \omega_{m}^{1-\delta - \alpha} - \lambda_{n}^{1 -\delta - \alpha} < \lambda_{m+1}^{1-\delta - \alpha} - \lambda_{n}^{1 -\delta - \alpha}$\\

iii) $\frac{1}{\lambda_{m+1}^{1-\delta - \alpha} - \lambda_{n}^{1 -\delta - \alpha}} \leq \frac{1}{\omega_{m}^{1-\delta - \alpha} - \lambda_{n}^{1 -\delta - \alpha}}$\\

iv) $\lambda_{m}^{\delta - \alpha}\displaystyle{\sum_{n=1}^{m}\frac{\mid\mid H_{\mu,\lambda}G^{-\delta}\mid\mid^{2}}{\lambda_{m+1}^{1 -\delta - \alpha} - \lambda_{n}^{1 -\delta - \alpha}} \leq \omega_{m}^{\delta - \alpha}}\displaystyle{\sum_{n=1}^{m}\frac{\mid\mid H_{\mu,\lambda}G^{-\delta}\mid\mid^{2}}{\omega_{m}^{1 -\delta - \alpha} - \lambda_{n}^{1 -\delta - \alpha}}}.$\\

With the aid of (2.25) the series $\displaystyle{\sum_{n=1}^{\infty}\frac{\mid\mid H_{\mu,\lambda}G^{-\delta}\mid\mid^{2}}{\omega_{m}^{1 -\delta - \alpha} - \lambda_{n}^{1 -\delta - \alpha}} \leq Const}.$\\ then\\

v)$\displaystyle{\mid\frac{1}{2i\pi}}\displaystyle{\int_{\gamma_{m}}Tr(H_{\mu,\lambda}R_{\sigma}^{0})^{2}})d\sigma \mid = o(\omega_{m}^{\delta - \alpha})$ $\hfill { }  (3.25)$\\

vi) The condition on the number $l$ of corrections necessary for the existence of finite formula of the trace of our operator is  $\delta -\alpha(l-1) \leq 0$ \\
i.e.\\
$ l \geq \frac{\delta}{\alpha} + 1$.$\hfill { }  (3.26)$\\

where $\frac{1}{2}\leq \delta < \frac{2}{3}$ and $0 \leq \alpha < \frac{2}{3} - \delta $ \\

vii) The minimal value of $l$ is obtained as follows:\\

 As $\frac{1}{2}\leq \delta < \frac{2}{3}$ then the condition (3.26) involves $ l \geq \frac{1}{2\alpha} + 1$ and  $\alpha < \frac{2}{3} - \frac{1}{2} = \frac{1}{6}$ i.e. $ 6 <\frac{1}{\alpha} $ and by (3.26) we deduce that $ l > 3 + 1 = 4 .$ therefore $l = 5$.}\\

{\bf Conclusion }\\

In {\bf[16]}, we study the trace of the semigroup $e^{-tH}$ where \\

 $ H = H_{\lambda{''},0,\mu,\lambda} = \lambda{''}a^{*3}a^{3} + \mu a^{*}a +i\lambda a^{*}( a + a^{*})a$\\
 
 We use the estimates obtained in {\bf[15]} which give an approximation of this semigroup by the unperturbed semigroup $ e^{-t\lambda''a^{*3}a^{3}}$ in nuclear norm. In particular to give an asymptotic expansion of this trace as $t \rightarrow 0^{+} $.\\

 This work is concluded  by noting that, if $\lambda{'} \neq 0$ and $\lambda \neq 0$, the existence of finite formula of the trace for the generalized Gribov operator $H_{\lambda'',\lambda',\mu,\lambda} = \lambda''a^{*3}a^{3} + \lambda'a^{*2}a^{2} + \mu a^{*}a + i\lambda a^{*}(a + a^{*})a$\\ is an open problem.\\

\begin{center}
{\bf References}
\end{center}

\begin{flushleft}
{\bf [1]} M.T. Aimar, A. Intissar and A. Jeribi , On an Unconditional Basis of Generalized Eigenvectors
of the Nonself-adjoint Gribov Operator in Bargmann Space, Journal of Mathematical Analysis and Applications 231, (1999), 588-602.\\
\end{flushleft}

\begin{flushleft}
{\bf [2]} T. Ando and M. Zerner, Sur une valeur propre d'un op\'erateur, Commun. Math. Phys. 93
(1984), 123-139\\
\end{flushleft}

\begin{flushleft}
{\bf [3]} V. Bargmann, On a Hilbert space of analytic functions and an associated integral
transform I, Commun. Pure Appl. Math. 14 (1962)\\
\end{flushleft}

\begin{flushleft}
{\bf[4]} L.A. Dikii, About a formula of Gelfand-Levitan, Usp. Mat. Nauk 8(2), (1953), 119-123.\\
\end{flushleft}

\begin{flushleft}
{\bf[5]} L.A. Dikii, New method of computing approximate eigenvalues of the Sturm-Liouville problem, Dokl. Akad. Nauk SSSR 116, (1957),  12-14.\\
\end{flushleft}

\begin{flushleft}
{\bf [6]} I.M. Gelfand and B.M.Levitan, On a Simple Identity for the Characteristic Values of a Differential Operator of Second Order, Dokl. Akad. Nauk SSSR, vol. 88, (1953), pp. 593-596\\
\end{flushleft}

\begin{flushleft}
{\bf [7]} I. Gohberg, S. Goldberg, and N. Krupnik, Traces and determinants of linear operators,
Birkhauser, 2000.\\
\end{flushleft}

\begin{flushleft}
{\bf [8]} I. C. Gohberg and M. G. Krein, Introduction to the Theory of Linear Non-Self Adjoint
Operators, Vol. 18, Am. Math. Soc., Providence, RI, (1969).\\
\end{flushleft}

\begin{flushleft}
{\bf [9]} V. Gribov, A reggeon diagram technique, Soviet Phys. JETP 26, no. 2, (1968), 414-423 \\
\end{flushleft}

\begin{flushleft}
{\bf [10]} A. Intissar, Etude spectrale d'une famille d'op$\acute{e}$rateurs non-sym$\acute{e}$triques intervenant dans la th$\acute{e}$orie des champs de reggeons, Commun. Math. Phys. 113 (1987), 263-297.\\
\end{flushleft}

\begin{flushleft}
{\bf [11]} A. Intissar, Quelques nouvelles propri$\acute{e}$t$\acute{e}$s spectrales de l'hamiltonien de la th$\acute{e}$orie des champs de reggeons, C.R. Acad. Sci. Paris, t. 308, Sér. I (1989), 209-214\\
\end{flushleft}

\begin{flushleft}
{\bf [12]} A. Intissar, Analyse Fonctionnelle et Th$\acute{e}$orie Spectrale pour les Op$\acute{e}$rateurs Compacts Non Auto-Adjoints, Editions Cepadues, Toulouse, (1997).\\
\end{flushleft}

\begin{flushleft}
{\bf [13]} A. Intissar, Analyse de Scattering d'un op$\acute{e}$rateur cubique de Heun dans l'espace de
Bargmann, Commun. Math. Phys. 199 ,(1998), 243-256.\\
\end{flushleft}

\begin{flushleft}
{\bf [14]} A. Intissar, M. Le Bellac, M. Zerner, Properties of the Hamiltonian of Reggeon field theory, Phys. Lett. B 113 (1982) 487-489.\\
\end{flushleft}

\begin{flushleft}
{\bf [15]} A. Intissar, Approximation of the semigroup generated by the Hamiltonian of Reggeon field theory
in Bargmann space, Journal of Mathematical Analysis and Applications, vol. 305, no. 2, ,
(2005), pp. 669-689\\
\end{flushleft}

\begin{flushleft}
{\bf [16]} A. Intissar, On regularized trace formula of Gribov semigroup generated by the Hamiltonian of Reggeon field theory in Bargmann representation, (2013) e-print.\\
\end{flushleft}

\begin{flushleft}
{\bf [17]} T. Kato, Perturbation Theory for Linear Operators (Springer-Verlag, New York, (1976).\\
\end{flushleft}

\begin{flushleft}
{\bf [18]} V. B. Lidskii, "Non-self-adjoint operators with a trace," Dokl. Akad. Nauk SSSR, 125,
No. 3, (1959), 485-487\\
\end{flushleft}

\begin{flushleft}
{\bf [19]} Naymark, MA: Linear Differential Operators. Nauka, M. 528 (1969)\\
\end{flushleft}

\begin{flushleft}
{\bf [20]} M. Reed and B. Simon, Methods of modern mathematical physics I: Functional analysis,
Academic Press, 1980.\\
\end{flushleft}

\begin{flushleft}
{\bf [21]} V.A. Sadovnichii, V.E. Podolskii,  On the class of Sturm-Liouville operators and approximate calculation of first eigenvalues, Mat Sbornik. 189(1), (1998), 133-148\\
\end{flushleft}

\begin{flushleft}
{\bf [22]} V.A. Sadovnichii and V.E. Podolskii, Traces of operators with relatively compact perturbations. Mat. Sb. 193 (2), (2002) 129-152\\
\end{flushleft}

\begin{flushleft}
{\bf [23]} V.A. Sadovnichii, V.E. Podolskii, Trace of operators. Uspech Math Nauk. 61(5), (2006), 89-156\\
\end{flushleft}

\begin{flushleft}
{\bf [24]} V. A. Sadovnichii and V. E. Podol'skii, Traces of Differential Operators, Differential Equations, Vol. 45, No. 4,(2009),  pp. 477-493.\\
\end{flushleft}

\begin{flushleft}
{\bf [25]} V. A. Sadovnichii and V. E. Podol'skii, Regularized Traces of Discrete Operators, Proceedings of the Steklov Institute of Mathematics, Pleiades Publishing, Inc.Suppl. 2, (2006),  pp. 161-177.\\
\end{flushleft}

\begin{flushleft}
{\bf [26]} B. Simon, Notes on infinite determinants of Hilbert space operators, Advances in Mathematics
24 (1977), pp. 244-273.\\
\end{flushleft}

\begin{flushleft}
{\bf [27]} B. Simon, Trace ideals and their applications, Mathematical Surveys and Monographs,
Volume 120, AMS, 2nd Ed. , (2005).\\
\end{flushleft}

\begin{flushleft}
{\bf [28]} N. G. Tomin, Several Formulas for the First Regularized Trace of Discrete Operators, Mathematical Notes, vol. 70, no. 1, (2001), pp. 97-109.\\
\end{flushleft}

 \end{document}